\documentclass[slac_one]{revtex4}
\usepackage{graphicx}
\usepackage{fancyhdr}
\newcommand{\jpsi}{{\ensuremath{J/\psi}}}
\newcommand{\ups}{{\ensuremath{\Upsilon(1S)}}}

\newcommand{\ptjpsi}{{\ensuremath{p_{T}^{J/\psi}}}}

\newcommand{\etajpsi}{{\ensuremath{\eta^{J/\psi}}}}

\begin{document}

%Title of paper
\title{Quarkonium Production Studies in CMS} %% Paper title goes here

\author{Z.~YANG$^{1,2}$ A.C.~KRAAN$^{2}$}

%by the European Union }}$}

\affiliation{1) SKLNST, Peking University,\ China \hspace{2mm} 2) Istituto Nazionale di Fisica Nucleare, Pisa, Italy}

\begin{abstract}
\vspace*{-0.4cm}
When the LHC starts its operation, CMS will have a unique opportunity to study quarkonium production in $pp$ collisions and later in $PbPb$ collisions. Here we report on the methods and plans for measuring the differential $p_T$~ \jpsi$\rightarrow\mu^+\mu^-$ production cross section, using data to be collected in the first LHC run by the CMS detector. Furthermore we discuss the performance of the CMS detector for quarkonium measurements in $PbPb$ collisions.
\end{abstract}

\maketitle

\thispagestyle{fancy}

\vspace*{-0.4cm}
\section{INTRODUCTION} % Section title should be in all capitals.
There are several reasons for studying quarkonia in $pp$ and $PbPb$ collisions at CMS. The main interest in quarkonia produced in $pp$ collisions arises from the fact that prompt quarkonium production continues to be puzzling: none of the theory models for quarkonium production at hadron colliders explains successfully both polarization and differential cross section measurements. In fact, the colour-octet mechanism~\cite{here}, part of the NRQCD approach, has been able to explain the inclusive quarkonium cross section data at the Tevatron, but predicts a polarization, which is in contrast with recent measurements at the Tevatron~\cite{cdf}. The colour singlet model (CSM) has recently been revived~\cite{maltoni}, but whether it can explain both cross section and polarization data is still uncertain. For reviews on quarkonium production see~\cite{here}. Thanks to the much higher collision energy and luminosity of the LHC, quarkonia can be probed with transverse momenta much higher than currently studied at the Tevatron, allowing us to discriminate better between theoretical models.

For $PbPb$ collisions in CMS, quarkonia are an excellent observable, both to study the thermodynamical properties of the quark-gluon plasma (QGP), as well as to constrain the parton distribution functions (PDFs) of the colliding nuclei. On the one hand, the higher centre-of-mass energy of the LHC with respect to RHIC will result in a higher temperature of the QGP, allowing us to study not only the suppression of \jpsi's, but now also that of $\Upsilon$'s. On the other hand, given that quarkonia are produced via gluon-gluon processes, the higher centre-of-mass energy at LHC will allow us to probe the gluon density of the nuclei at much lower parton fractional momenta than previously studied. 
 
%Three processes dominate \jpsi~ production in $pp$ and $PbPb$ collisions: prompt \jpsi's produced directly, prompt \jpsi's produced indirectly (via decay of excited states), and non-prompt \jpsi's from the decay of a B-hadron. 
%It is the prompt production of quarkonia which continues to be particularly puzzling. 

%Prompt quarkonium production at hadron colliders has been studied extensively in the theory of NRQCD. One of the innovations of NRQCD is the so-called colour-octet mechanism (COM), which successfully explains the inclusive quarkonium cross section data at the Tevatron. However, recent polarization measurements, which revealed unpolarized \jpsi's, are in clear contrast with COM predictions. The colour singlet model (CSM) has recently been revived~\cite{maltoni}, but whether it can explain both cross section and polarization data is still uncertain. For reviews on quarkonium production see~\cite{here}. 

CMS can contribute to the study of quarkonia thanks to its excellent capabilities to detect muons from quarkonium decays up to large pseudorapidity ($|\eta|<2.5$). Given the large yields of quarkonia which will be produced at CMS, many analyses dedicated to quarkonium production are viable already with small integrated luminosity.
%Thanks to the much higher collision energy and luminosity both in $pp$ and $PbPb$ collisions, quarkonia can eventually be probed in different regions of phase space than currently studied at the Tevatron and RHIC. 
Quarkonia are additionally relevant for CMS in the early phases of LHC running because they are crucial for detector alignment and calibration. 

In this note we describe a feasibility study to measure the \jpsi$\rightarrow\mu^+\mu^-$ differential cross section in $pp$ data at $\sqrt{s}$=14 TeV. In addition we discuss the performance for observing quarkonia in $PbPb$ collisions at $\sqrt{s_{NN}}=5.5$ TeV. 
\vspace*{-0.3cm}\section{J/\boldmath{$\psi\rightarrow\mu^+\mu^-$} DIFFERENTIAL CROSS SECTION IN \boldmath{$PP$} COLLISIONS}
%\vspace*{-0.3cm}
Three processes dominate \jpsi~hadroproduction: prompt \jpsi's produced directly, prompt \jpsi's produced indirectly (via decay of excited states), and non-prompt \jpsi's from the decay of a B-hadron. CMS proposes the following study to measure the inclusive, prompt, and non-prompt contributions to the differential cross section.

%\subsection{Quarkonium triggers in CMS}
%The LHC bunch crossing rate at nominal luminosity will be 40 MHz. Via a Level 1 (L1) and high level trigger (HLT) system the rate will be reduced to 100 Hz before the events are stored. There are two types of triggers relevant for quarkonia. The most important trigger, which is used in this analysis, is the dimuon trigger with a threshold for both muons of 3 GeV/c. employing fast \jpsi~reconstruction. 
%HLT mass windows of XXX-XXX for selecting \jpsi's and $\psi(2S)$, and YYY-YYY for selecting \ups, \upss and \upsss). The L1 and HLT trigger muon $p_T$ threshold for both muons is 3 GeV/c (the $2\mu 3$ menu).
% and the menu is referred to as $2\mu 3$. 
%In addition there is a HLT menu based on selecting events with a displaced dimuon vertex, for collecting for example non-prompt \jpsi-events.
%, such as $B_s\rightarrow\jpsi\phi$ and $B^+\rightarrow\jpsi K$, and also non-resonant dimuon events such as $B_s\rightarrow \mu^+\mu^-$.
%\vspace*{-0.12cm}
%In addition single muon triggers may be used.
%, where both at L1 and at HLT one muon is required with a certain $p_T$. 
%MAYBE RATES? NO TRIGGER FIGURES.
%These may be usDifferent trigger thresholds are anticipated and the lowest thresholds will be prescaled triggers.
%Depending on luminosoty these menues may be prescaled.
\subsection{\jpsi~reconstruction}
For this study~\cite{cmspas} prompt \jpsi~signal events were generated with PYTHIA 6.409, including direct and indirect production. Non-prompt \jpsi~events were obtained by generating minimum bias events with PYTHIA. 
%In both cases the presence of 2 muons with \ptmu$>2.5$ GeV/c and $|\etamu|<2.5$ was required at generator level. 
As background events were considered any other source of muons that, when paired, could accidentally have an invariant mass close to that of the \jpsi. These muons come mainly from heavy flavoured quark decays and sometimes from a $\pi^{\pm}$ or $K^{\pm}$ decay in flight. The amount of background from Drell-Yan events was found to be below 1\% with respect to that of other background sources.

%The following sources were considered:
%\begin{itemize}
%\item generic QCD 2$\to$2 events produced in PYTHIA, requiring the presence of one $\mu$
%% with $\ptmu>2.5$ GeV/c and $|\etamu|<2.5$ 
%at generator level, mainly coming from heavy flavoured quark decays. 
%%These events are referred to as ''muon enriched QCD background" 
%%in the following. 
%Although this sample does include combinatons of one muon from heavy quark decay and one muon from a $\pi^{\pm}$ or $K^{\pm}$ decay in flight,  background from two muons from decay-in-flight is not included. The effect of omitting this background is small and included in the systematic uncertainties.
%\item Drell-Yan events.
%% where both muons have  $\ptmu>2.0$ GeV/c and $|\etamu|<2.5$.
%\end{itemize}

\begin{figure}[h!]
\centering
\includegraphics[width=7.5cm,height=7.cm]{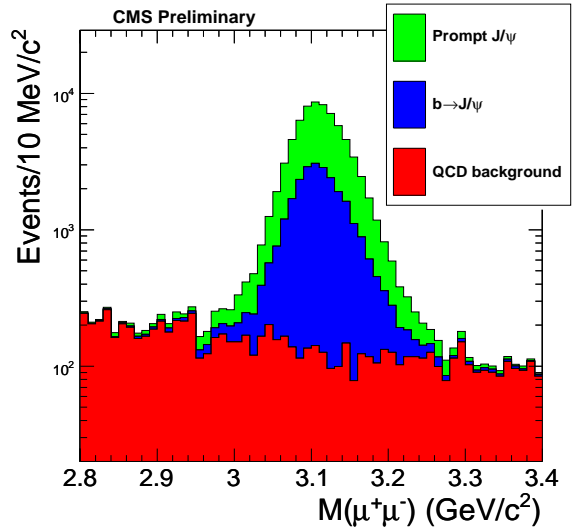}
\caption{Dimuon mass distribution normalized to
3~pb$^{-1}$ in logarithmic scale. The green (light grey), blue (black) and red (dark grey) areas are the prompt, non-prompt and background contributions, respectively. 
%Right: the $\ell_{xy}$ distribution for  prompt and non-prompt \jpsi's. 
}\label{22}
\end{figure}

\jpsi~events are selected with the dimuon trigger employing fast \jpsi~reconstruction with a transverse momentum threshold for both muons of at least 3 GeV/c~\cite{cmstdr}. \jpsi~candidates are then reconstructed by pairing muons with opposite charge. The invariant mass of the muon pair is required to be between 2.8 and 3.4 GeV/c$^2$. The two muons are required to come from a common vertex. The dimuon mass spectrum including background and signal is given in Fig.~\ref{22}. About 75000 \jpsi's are expected to be reconstructed in 3 pb$^{-1}$. The mass resolution is about 30 MeV/c$^2$. The distribution is fitted fit with a double Gaussian for the \jpsi~signal and a linear function for background.
The resulting number of fitted events for the signal, divided by the
total number of generated events, defines the signal reconstruction
efficiency convoluted with acceptance, which depends on \ptjpsi,
\etajpsi~and on the \jpsi~polarization. Below \ptjpsi$\sim 5$ GeV/c the reconstruction efficiency vanishes. At \ptjpsi$>20$ GeV/c it is about 60\% for unpolarized \jpsi. Effects of the polarization are taken into account in the systematic uncertainties.

\subsection{Measurement of inclusive, prompt and non-prompt \jpsi~cross section}
The inclusive \ptjpsi~ differential cross section  measurement, covering
the region $|\etajpsi|<$2.4, is based on the following expression:
\begin{equation}
\frac{d\sigma}{dp_{T}}(J/\psi)\cdot{}Br(J/\psi\rightarrow \mu
^{+}\mu^{-})=\frac{N^{fit}_{\jpsi}}{\int{}
Ldt\cdot{}A\cdot \Delta p_{T}}
\end{equation}
where $N^{fit}_{\jpsi}$ is the number of reconstructed $J/\psi$ in a given $p_{T}$ bin resulting from the mass spectrum fit as explained above, $\Delta p_{T}$  is the size of the $p_{T}$ bin, $\int{}Ldt$ is the integrated luminosity, and $A$ is the total efficiency taking into account trigger and reconstruction efficiencies.
% determined by comparing the measured distributions with those simulated by Monte Carlo. CMS will make use of the so called ''tag and probe" methods  to determine all corrections.
%~\cite{tag_probe}. 
%Fig.~\ref{figxs} displays the expected inclusive \jpsi~cross section for 3 pb$^{-1}$ of data.

\begin{figure}[t!]
\centering
%\includegraphics[width=7cm,height=7cm]{Prompt_ct.pdf}
 %\hspace{1cm}
%\includegraphics[width=7cm,height=7cm]{Bdecay_ct.pdf}
 %\hspace{1cm}
\includegraphics[width=7.cm,height=7.cm] {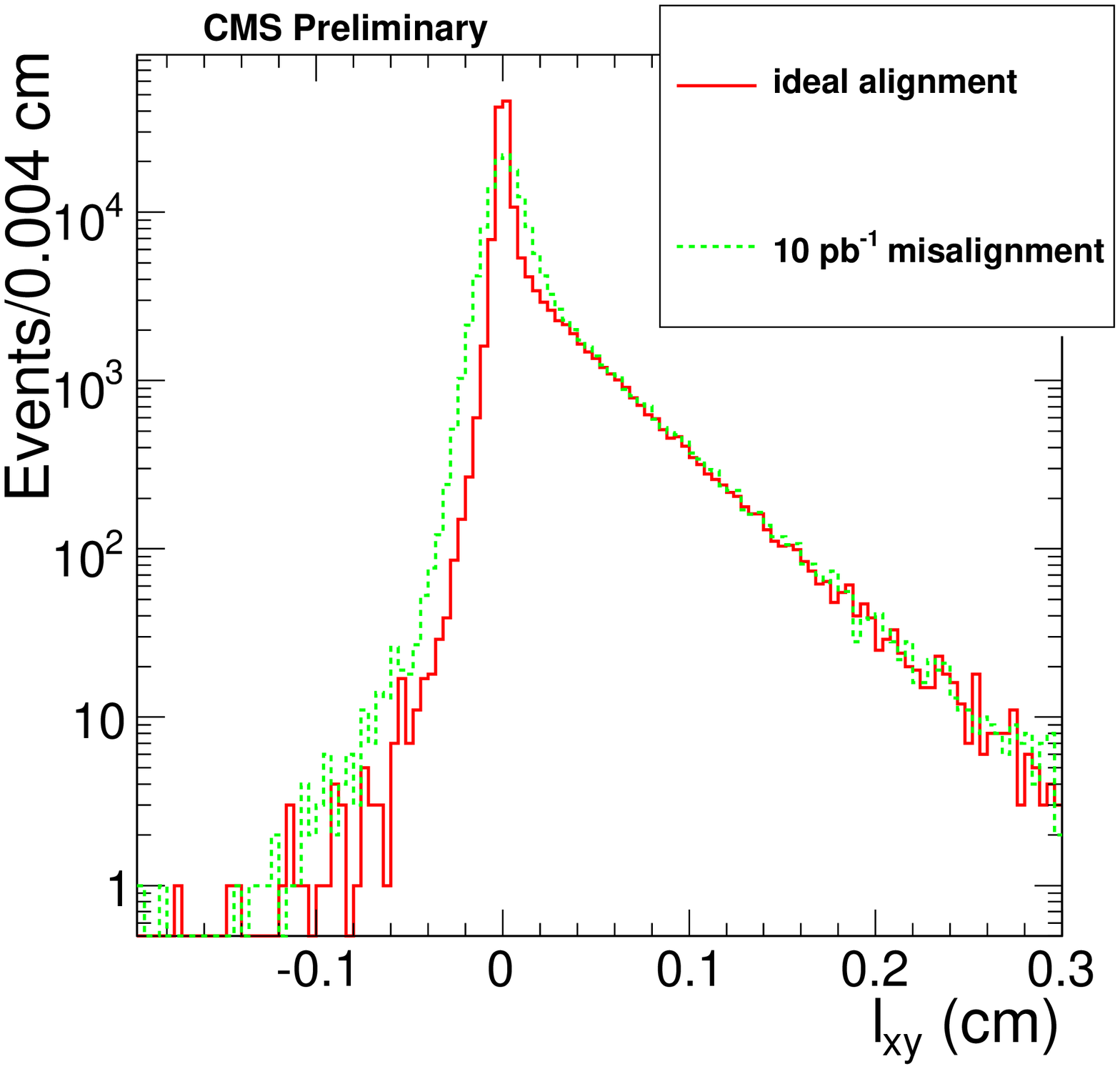}
\includegraphics[width=7.cm,height=7.cm]{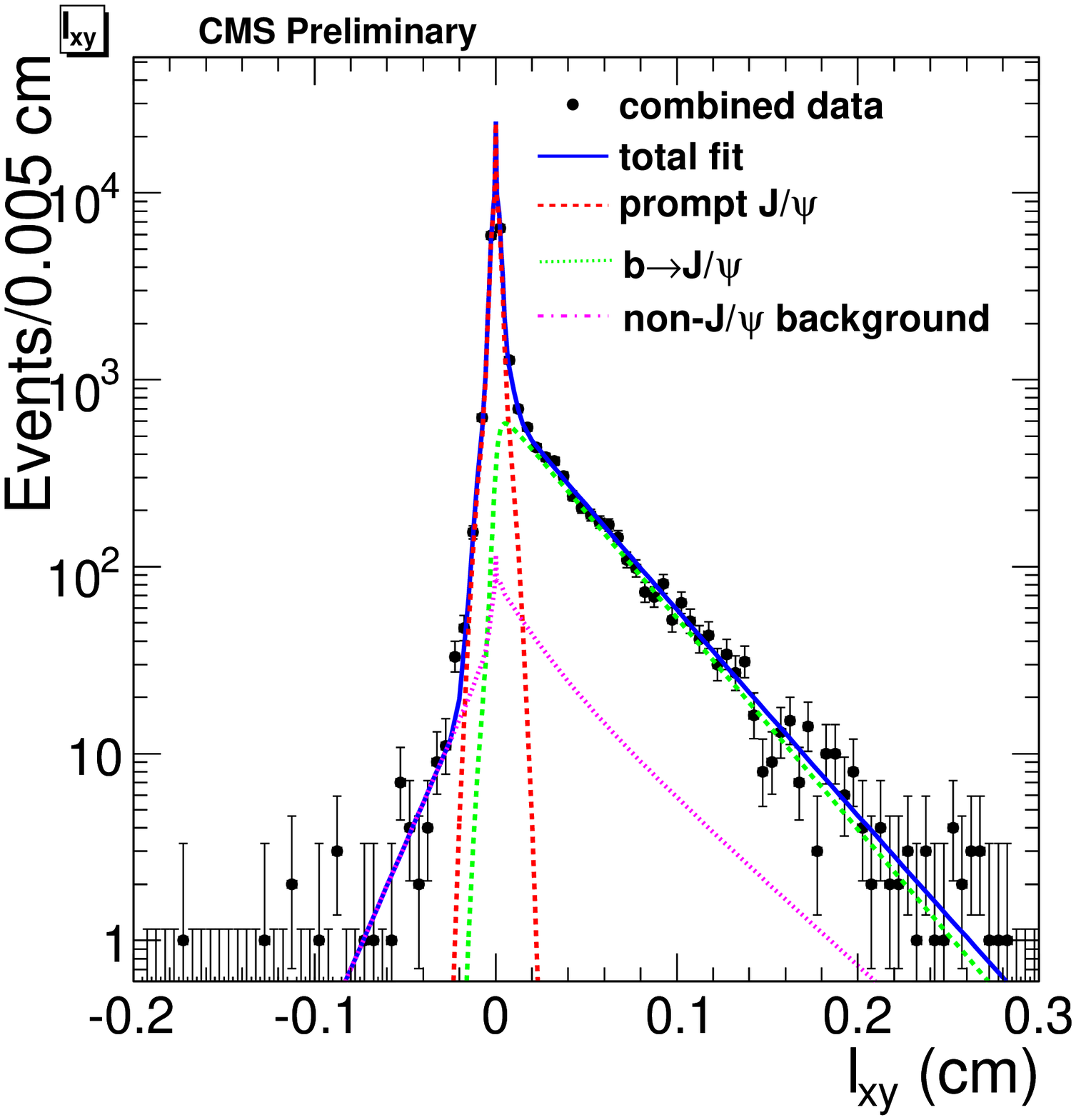}
%Combined_ct_5.pdf}
\vspace*{-0.3cm}
\caption{Left : the $\ell_{xy}$ distribution for  prompt and non-prompt \jpsi's. Right: distribution of $\ell_{xy}$ and likelihood fit result in the range of
9 $<p_{T}<$10 GeV/$c$. }\label{34}
\end{figure}

%\subsection{Disentangling prompt and non-prompt \jpsi's}
For each \jpsi~candidate $\ell_{xy}=L_{xy}\cdot m_{\jpsi}/{\ptjpsi}$ is computed,
where $L_{xy}$ is the distance in the transverse plane between the  vertex of the two muons and the primary vertex of the event. The $\ell_{xy}$ distribution for prompt and non-prompt $J/\psi$'s is shown in Fig.~\ref{34} (left). For prompt \jpsi's the $\ell_{xy}$ distribution is given by a resolution function, while for non-prompt \jpsi's a resolution function convoluted with an exponential is used. To determine the fraction  $f_B$ of $J/\psi$'s from B-hadron decays, an unbinned maximum-likelihood fit was performed fitting simultaneously the invariant mass and lifetime distributions. Figure~\ref{34} (right) shows an example of a fit for 9 $<p_T^\jpsi<$ 10 GeV/c.
 % The difference between the fitted fraction and the true Monte Carlo fraction can be seen in Fig.~\ref{35} (right).
\subsection{Results}
\begin{figure}[h!]
\vspace*{-0.3cm}
\centering
\includegraphics[width=5.8cm,height=5.8cm]{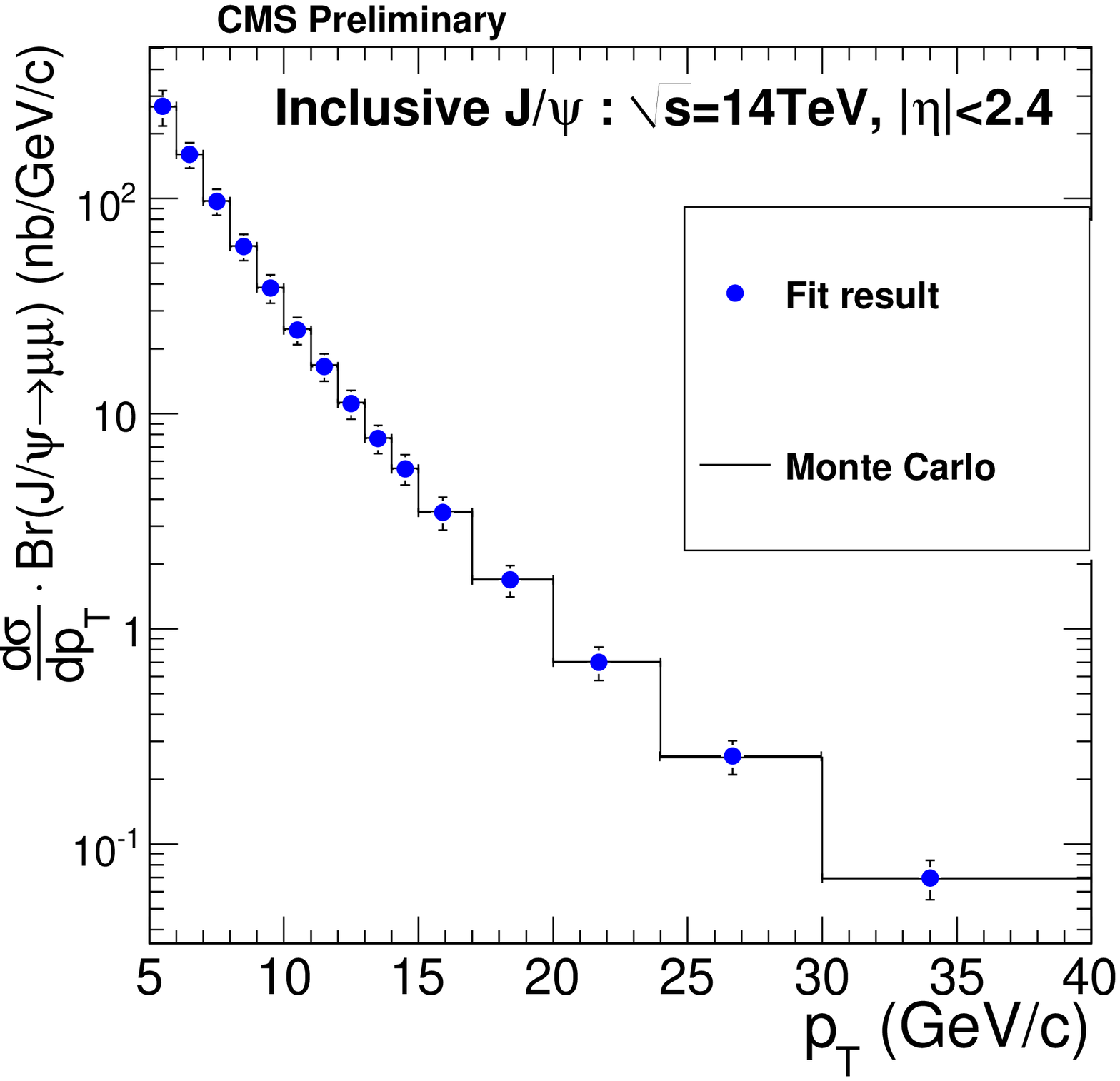}
\includegraphics[width=5.8cm,height=5.8cm]{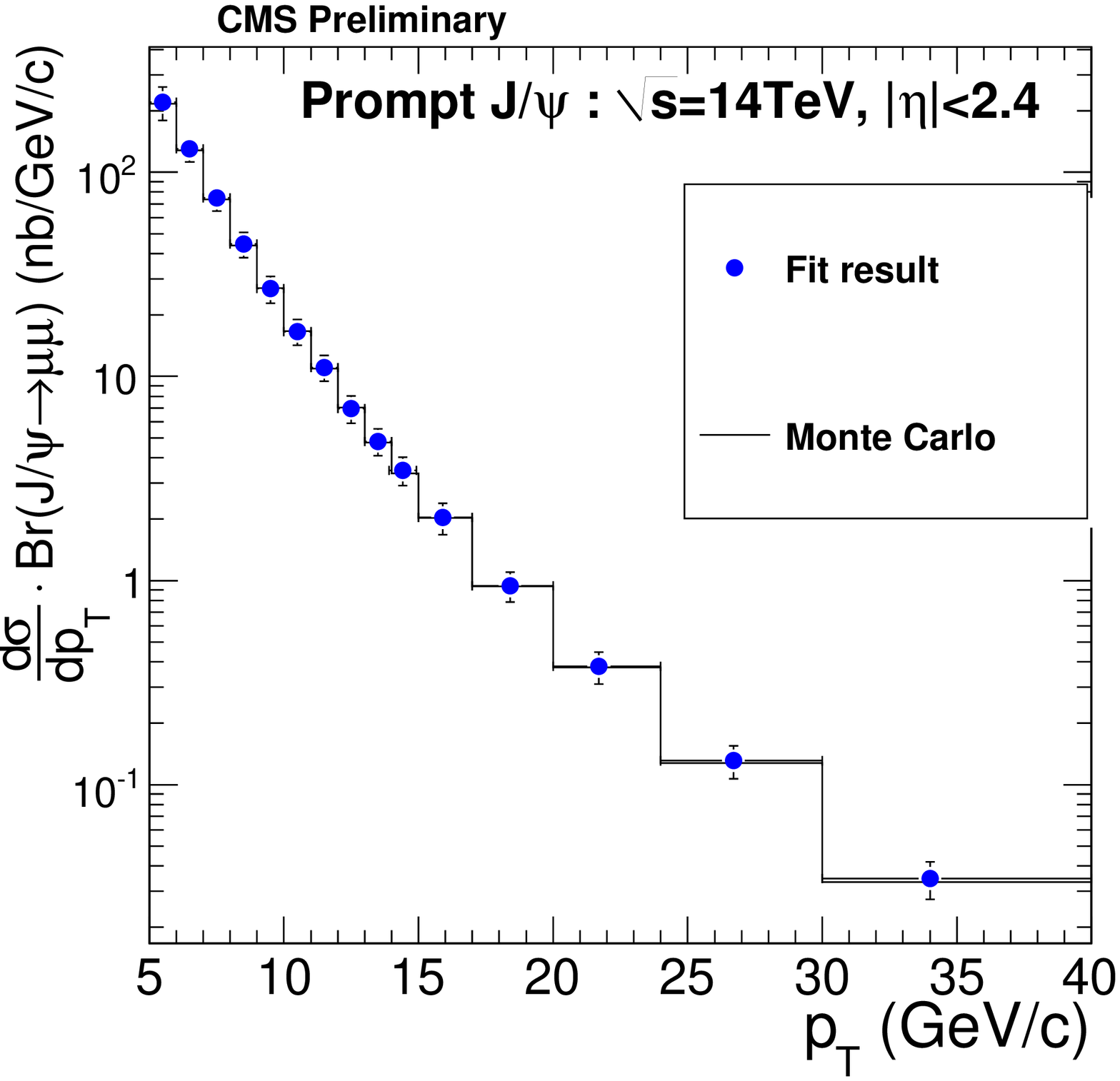}
\includegraphics[width=5.8cm,height=5.8cm]{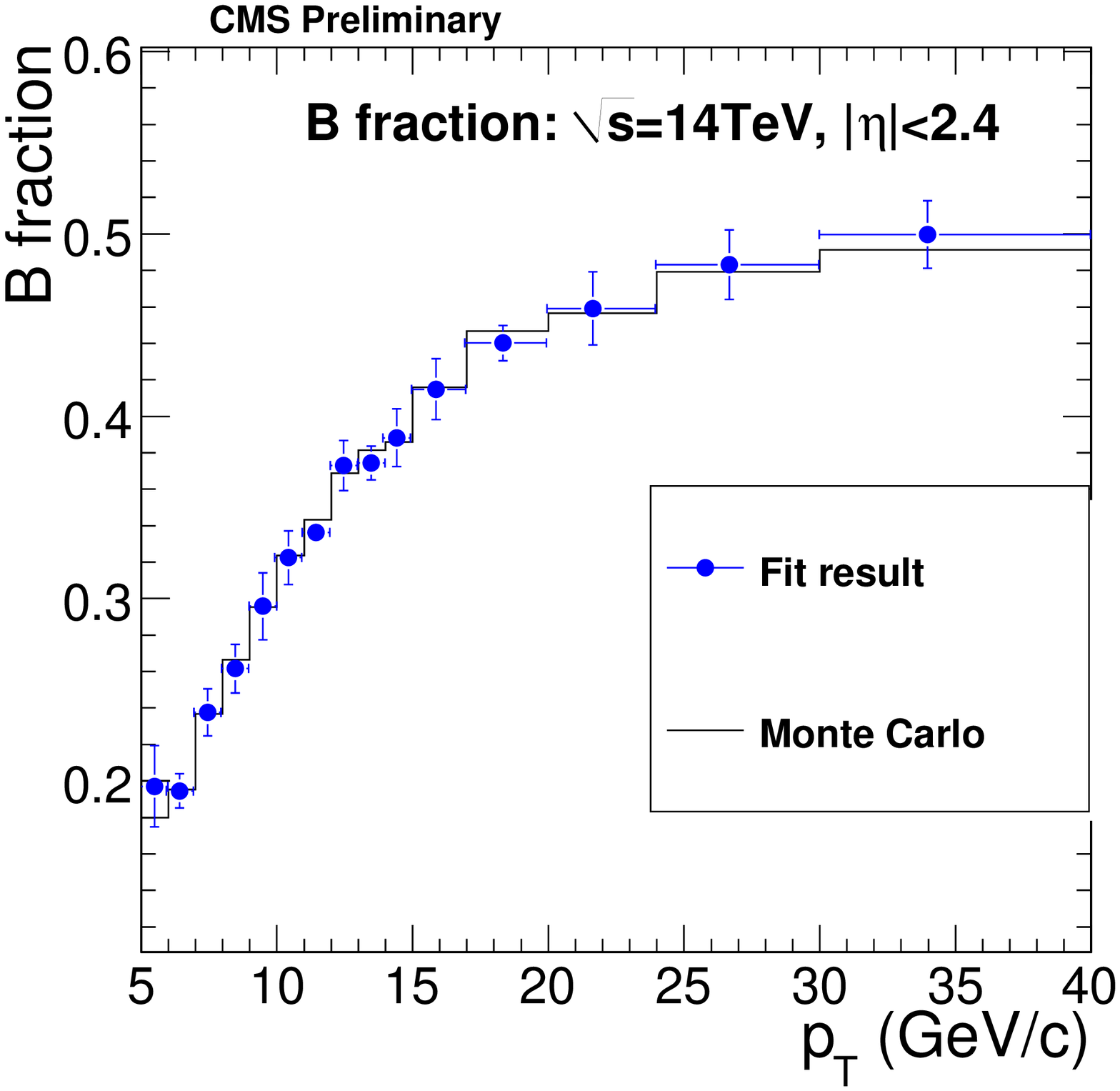}
\caption{The inclusive (left) and prompt (center) $J/\psi$ differential cross section $d\sigma/dp_{T}\cdot{}Br(J/\psi\rightarrow \mu ^{+}\mu^{-})$ and the fitted fraction $f_B$ of \jpsi's from B-hadron decays (right), as a
function of \ptjpsi, integrated over $|\etajpsi|<2.4$ for an integrated luminosity of 3~pb$^{-1}$. } \label{81}
\end{figure}

The dominant systematic uncertainties are the luminosity and the \jpsi-polarization. The total uncertainty is about 13\% at \ptjpsi$>20$ GeV/c and 19\% in the lowest \ptjpsi~bin, from 5-6 GeV/c. The inclusive and prompt \jpsi$\rightarrow\mu^+\mu^-$~cross sections are given in Fig.~\ref{81} left and center respectively. The fraction $f_B$ of \jpsi's from B-hadron decay is given in Fig.~\ref{81} (right).
\newpage
\section{QUARKONIUM MEASUREMENTS IN HEAVY-ION COLLISIONS}
\begin{figure}[h!]
\vspace*{-0.3cm}
\centering
\includegraphics[width=6.9cm,height=5.7cm]{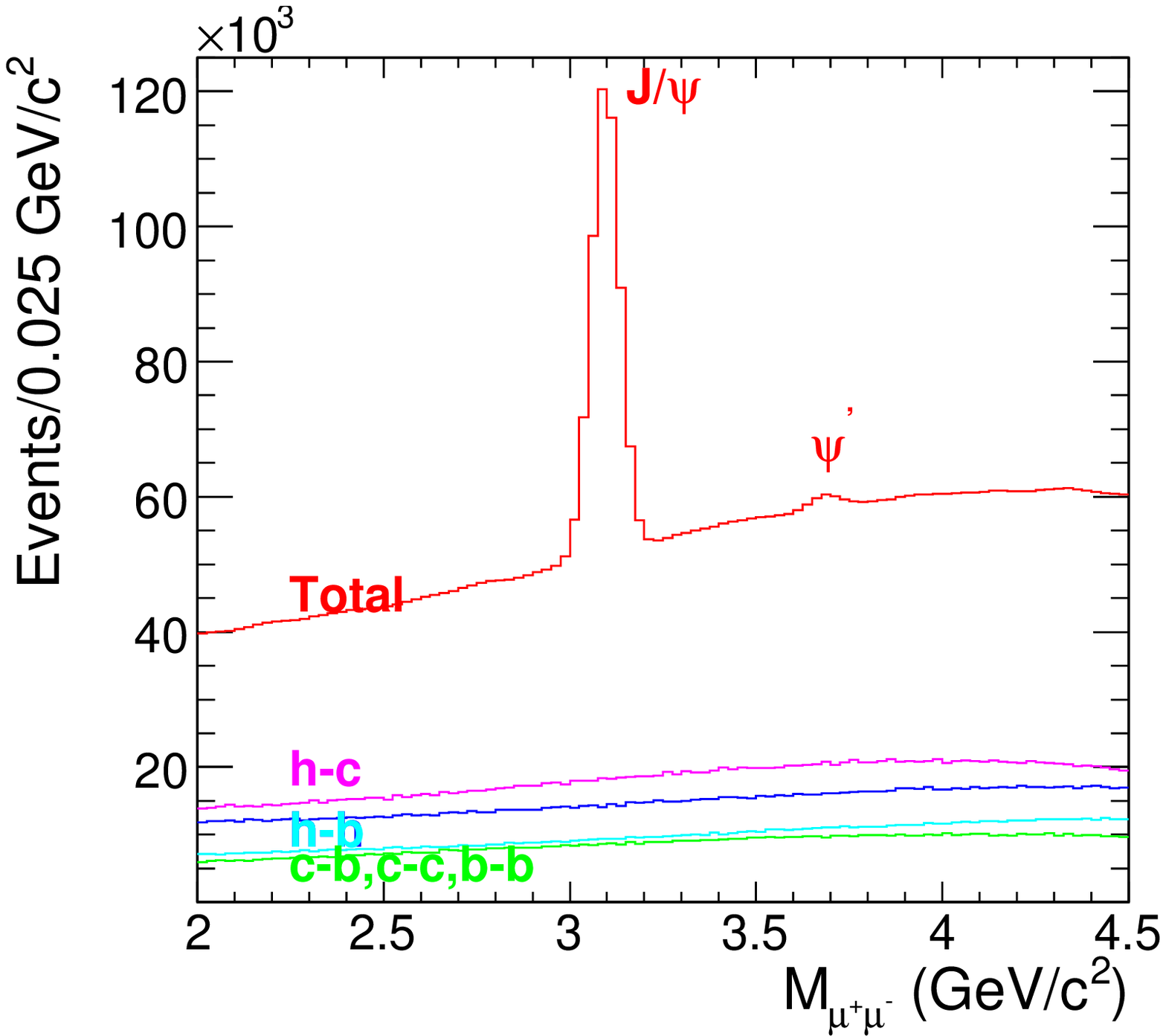}
\includegraphics[width=6.9cm,height=5.7cm]{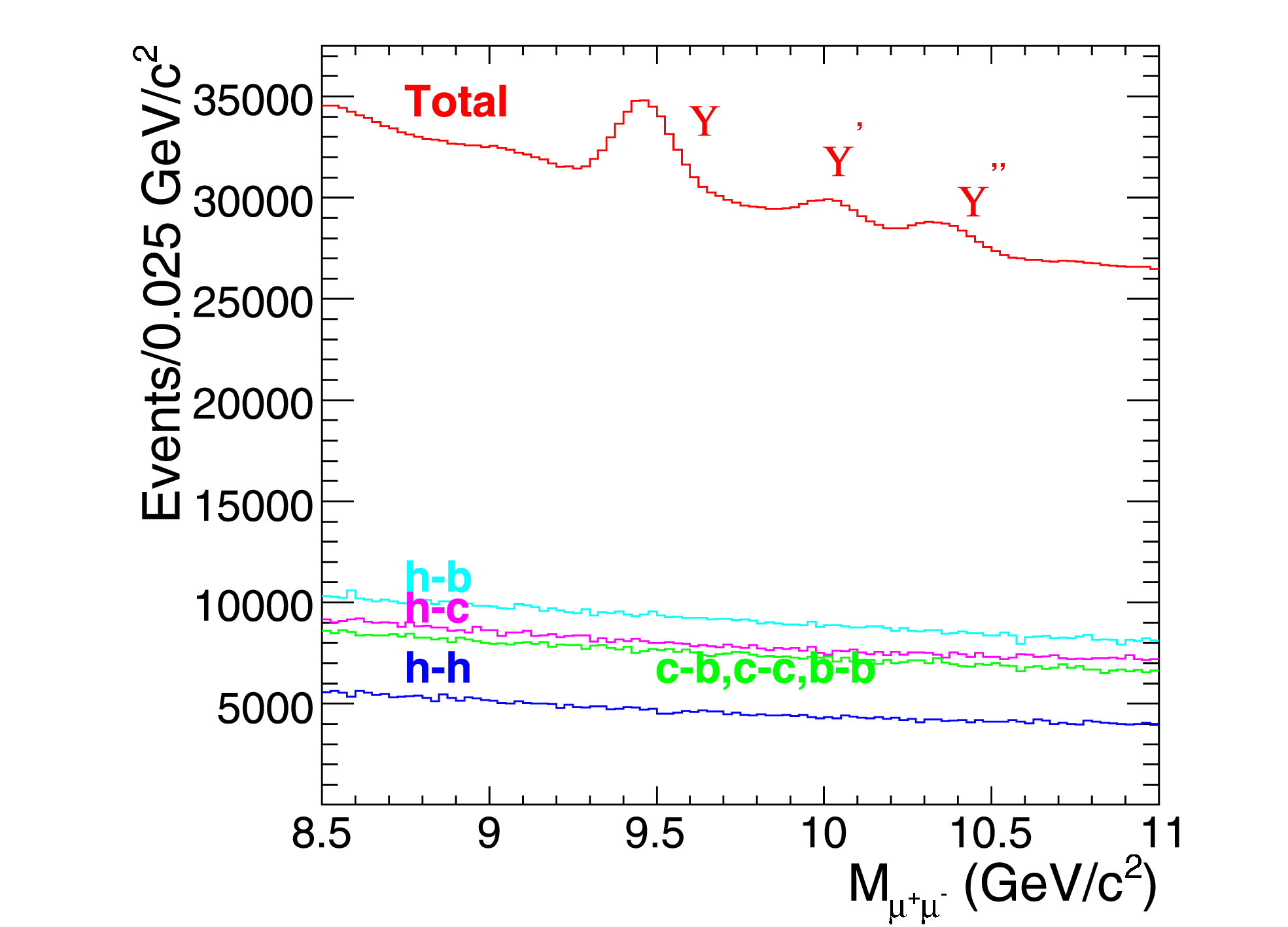}
\vspace*{-0.3cm}
\caption{Dimuon mass distributions measured within $|\eta|<2.4$ for 0.5 nb$^{-1}$ of $PbPb$ data with d$N_{ch}/$d$\eta_{\eta=0}$=2500 in the \jpsi~(left) and $\Upsilon$~(right) mass regions, together with the background. } \label{hi}
\end{figure}

CMS has also studied the performance for reconstructing quarkonia in a $PbPb$ collision environment~\cite{cmstdr_hi}. 
%The performance is expected to be very similar to that in $pp$ collisions. 
Fig~\ref{hi} shows the dimuon mass distributions in the \jpsi~(left) and $\Upsilon$~(right) mass regions corresponding to 0.5 nb$^{-1}$ of $PbPb$ data. The muon and \jpsi~reconstruction performance in $PbPb$ collisions is very similar to that in $pp$ collisions, hence the mass resolution for \jpsi's and \ups~is roughly equal to what is expected in $pp$ collisions. What differs is the amount of background as well as the quarkonium yield. Significantly more background is expected from combinatorial background of various muon sources (pion, kaon and heavy-quark decays) which can be removed using techniques such as like-sign subtraction. In 0.5 nb$^{-1}$ of data (about one month of $PbPb$ running) about 180000 \jpsi's and 25000 \ups's are expected to be collected. Such a large statistics sample will allow us to compare in detail central and peripheral $PbPb$ collisions and to carry out $p_T-$ and $y-$differential cross section studies which will contribute to clarify the physics mechanisms behind the production (and destruction) of quarkonium states in high-energy nucleus-nucleus collisions.

\vspace*{-0.8cm}
\section{CONCLUSION}
\vspace{-0.2cm}
We have presented quarkonium studies in early CMS $pp$ collisions and in $PbPb$ collisions. More quarkonium studies in CMS are ongoing. First results based on real data are expected in 2009. 
\vspace*{-0.5cm}

\begin{acknowledgments}
\vspace*{-0.1cm}
Thanks to David d'Enterria for discussions about heavy-ion physics and for comments on this note. Also thanks to Sijin Qian for comments on this note. This work is partially supported by National Natural Science Foundation 
of China (Contract No. 10099630), the Ministry of Science and 
Technology of China (Contract No.2007CB816101). Aafke Kraan is financed by the European Union as Marie Curie fellow (Contract No. EIF PHY-040156).
\end{acknowledgments}

\end{document}